\documentclass[12pt,a4paper,twoside]{article}
\usepackage{epsfig}
\usepackage{baltlat1}
\usepackage{wrapfig}
\begin{document}
\ \
\vspace{0.5mm}

\setcounter{page}{1}
\vspace{8mm}

\titlehead{Baltic Astronomy, vol.12, 604--609, 2003.}

\titleb{CORRECTION OF $UBV$ PHOTOMETRY \\
        FOR EMISSION LINES}

\begin{authorl}
\authorb{A.~Skopal}{}
\end{authorl}

\begin{addressl}
\addressb{}{Astronomical Institute, Slovak Academy of Sciences,
          059\,60 Tatransk\'a Lomnica, Slovakia}
\end{addressl}

\submitb{Received October 15, 2003}

\begin{abstract}
We investigated the effect on the $U,~B,~V$ magnitudes of
the removal of the emission lines from the spectra of some 
symbiotic stars and novae during their nebular phases. 
We approached this problem by a precise reconstruction of 
the composite UV/optical continuum and the line spectrum. 
Corrections $\Delta U$, $\Delta B$, $\Delta V$ are 
determined from the ratio of fluxes with and without 
emission lines. 
Here we demonstrate this effect on the case of the symbiotic nova 
V1016~Cyg during its nebular phase. We find that about 68\%, 78\% 
and 66\% of the observed flux in the $U$, $B$ and $V$ filters is 
radiated in the emission lines. 
The effect should be taken into account before using the observed 
color indices of emission-line objects for diagnosis of their 
radiation in the continumm.
\end{abstract}

\resthead{Correction of the UBV photometry for emission lines}{A.~Skopal}



\sectionb{1}{INTRODUCTION}

Photometric measurements in the standard $U,~B,~V$ filters are 
often used to analyze radiation in the continuum of many kinds 
of stellar objects. A diagnostic by the ($U-B,~B-V$)-diagram is 
frequently applied to compare the observed colour indices to 
those of the continuum radiation from main-sequence stars, 
supergiants, a blackbody and/or a nebula. However, the true 
continuum is often affected by the line spectrum, which thus 
requires corrections of the photometric observations before 
studying the continuum radiation. For example, the presence 
of emission lines in the spectral region of the $U,~B,~V$ 
passbands leads to {\em brighter} magnitudes than those of 
the continuum. 

In this contribution we introduce the effect on the $U,~B,~V$ 
magnitudes due to the removal of emission lines from the spectrum. 
However, a strong variation of the emission spectrum due to 
the activity, large differences between individual objects 
and a complex profile of the true continuum for emission-line 
objects preclude a simple solution. 
Hither-to, this problem has been approached only by a few 
groups of authors and without giving any concept for a general 
application (see, e.g., Fernandez-Castro et al. 1995)

Here we quantify corrections of the $U,~B,~V$ photometry for 
emission lines by exact calculations of the predicted spectrum. 
We demonstrate this effect on the case of symbiotic nova 
V1016~Cyg, which spectrum is very rich to numerous strong 
emission lines during its nebular phase. 

\sectionb{2}{ANALYSIS}

\subsectionb{2.1}{The method}

The aim of this paper requires the ratio of the continuum with
the superposed emission lines to the line-removed continuum at
all relevant wavelengths. Thus we need the profile of the 
continuum and the emission line spectrum obtained (in the ideal 
case) simultaneously with the photometric observations. 

To quantify the effect of emission lines on the $U,~B,~V$
measurements, we express the observed flux in the form
%
\begin{equation}
  F_{\rm obs}(\lambda) =
  F_{\rm cont}(\lambda) (1 + \epsilon (\lambda)),
\end{equation}
%
where $F_{\rm cont}(\lambda)$ is the true continuum (i.e. 
line-removed continuum) and $\epsilon (\lambda)$ represents 
the emission line spectrum in units of the continuum at the 
wavelength $\lambda$. Then the magnitude difference, $\Delta m$, 
between the observed magnitude, $m_{\rm obs}$, and the magnitude 
of the true continuum, $m_{\rm cont}$, can be expressed as 
\begin{eqnarray}
\Delta m = m_{\rm obs} - m_{\rm cont} = 
~~~~~~~~~~~~~~~~~~~~~~~~~~~~~~~~~~~~~~
\nonumber \\
  -2.5\log\,\biggl[\int_{\lambda}\!\! F_{\rm cont}(\lambda)S(\lambda)
               (1 + \epsilon (\lambda))\,{\rm d}\lambda \bigg/\!\!
               \int_{\lambda}\!\!F_{\rm cont}(\lambda)S(\lambda)\,
               {\rm d}\lambda \biggr], 
\end{eqnarray}
where $S(\lambda)$ are transmission functions of the $U,~B,~V$ 
filters. 
Further we approximate the emission line spectrum with an ensemble
of Gauss functions, $G_{\rm i}$, as
%
\begin{equation}
 \epsilon (\lambda) = \sum_{i}G_{\rm i}(\lambda;\lambda_{\rm i},
                      I_{\rm i},\sigma_{\rm i}),
\end{equation}
%
where $\lambda_{\rm i}$ is the wavelength of the i-$th$ line,
$I_{\rm i}$ its maximum in units of the local continuum and
$2\sigma_{\rm i}$ its FWHM. According to the relation (2) 
%
%
\begin{wrapfigure}{i}[0pt]{61mm}
\centerline{\psfig{figure=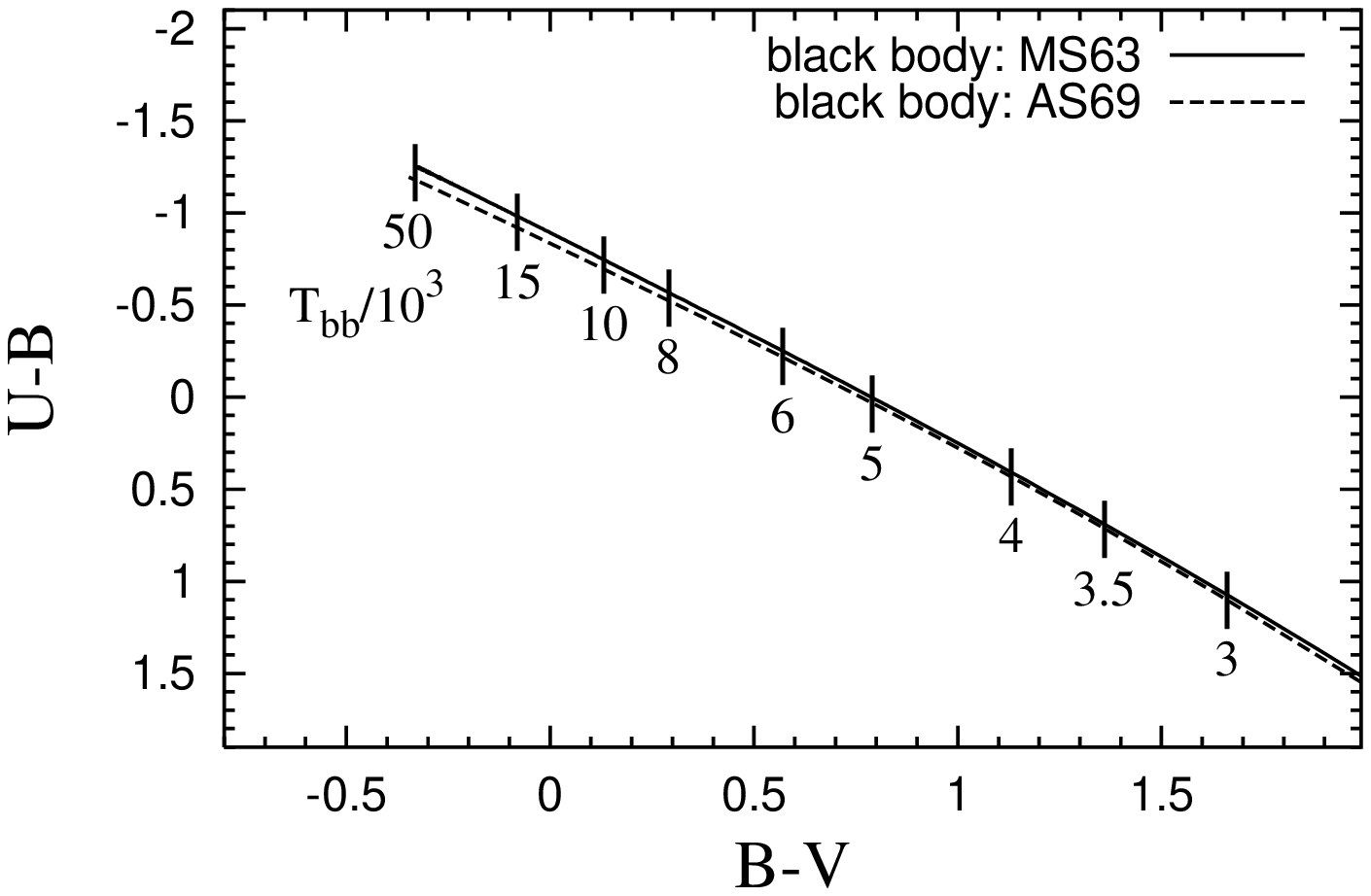,width=40truemm,angle=-90,clip=}}
\captionb{1}{Comparison~of~the~blackbody 
colour~indices~calculated~according~to 
Eq.~5~(solid~line, Matthews \& Sanda\-ge 1963) and Eq.~6~(dashed line, 
A\v{z}usienis \& Strai\v zys 1969). 
}
\end{wrapfigure}
%
the removal of emission lines from
the spectrum gives {\em fainter} magnitudes at all wavelengths. 
A main difficulty in calculating corrections $\Delta m$ is
connected with reconstruction of the continuum profile, 
$F_{\rm cont}(\lambda)$, which can be rather complex 
in the case of the composite continuum of symbiotic stars. 
We introduce briefly this problem in Sect. 2.3. 
We reconstructed the line spectrum, $\epsilon (\lambda)$, 
according to its parameters available in the literature 
(fluxes and the continuum level). 
Example is given below in Sect. 2.4. 

\subsectionb{2.2}{Calculated $U - B$ and $B - V$ colours}

The colours on the $U,~B,~V$ system can be determined once 
the transmission functions $S(\lambda)$ of the system are 
known. Theoretical colours $(U - B)_{0}$ and $(B - V)_{0}$ 
can be calculated as 
%
\begin{equation}
(U - B)_{0} =
  -2.5\log\,\biggl[\int_{\lambda}\!\! F(\lambda)S_{\rm U}(\lambda)\,
                {\rm d}\lambda
          \bigg/\!\!\int_{\lambda}\!\!F(\lambda)S_{\rm B}(\lambda)\,
                {\rm d}\lambda \biggr]
\end{equation}
%
with a similar equation for the $(B - V)_{0}$ index.
The aim is to obtain such $U - B$ and $B - V$ indices, which 
would predict the observed colours for real stars of known
energy distribution. This task requires additional colour 
equations, which are used to convert theoretical calculations 
based on the adopted $S(\lambda)$ to the empirical $U,~B,~V$ 
system. Matthews \& Sandage (1963) derived these equations as 
%
\begin{equation}
  B - V = 1.024 (B - V)_{0} + 0.81, ~~~~~~
  U - B = 0.921 (U - B)_{0} - 1.308, 
\end{equation}
which correspond to $S(\lambda)$ from their Table A1. 
Later on, A\v{z}usienis \& Strai\v zys (1969) 
suggested the following relations 
\begin{equation}
  B - V = (B - V)_{0} + 0.67, ~~~~~~
  U - B = (U - B)_{0} - 1.33,
\end{equation}
for their set of the revised response functions (see their Table~1). 
We compared both transformations on the example of the blackbody 
colour indices (Figure~1). Both transformations give very close 
values, and in this paper we use only 
the Matthews \& Sandage (1963) equations. 

\subsectionb{2.3}{The composite continuum}

We reconstruct the continuous radiation, $F_{\rm cont}(\lambda)$, 
of symbiotic binaries by a three-component model, which
consists of two stellar components of radiation -- from the hot
and the cool star -- and the nebular radiation from the ionized
circumbinary medium (e.g. Nussbaumer \& Vogel 1989). We approach 
this problem with the aid of low-resolution IUE spectra and 
the broad-band infrared photometry. The latter was 
approximated by synthetic spectra for red giants according 
to models of Hauschildt (1999) to get the infrared stellar 
continuum. Then, on the basis of such defined continuum, 
we applied the three-component model of radiation of symbiotic 
stars to get the profile of the continuum in between these 
regions (see Skopal 2001, 2003 for more details). 

\subsectionb{2.4}{The emission line spectrum}

Figure~2 shows example of the $\epsilon (\lambda)$ function 
from Eq.~1, which represents the emission line spectrum of the 
symbiotic nova V1016~Cyg. To reconstruct this function 
we used emission line fluxes published by Schmid \& Schild (1990) 
(hereafter SS90). The spectrum was taken on 15/11/87 at the INT 
telescope. 
\vskip1mm
\begin{figure}
\centerline{\psfig{figure=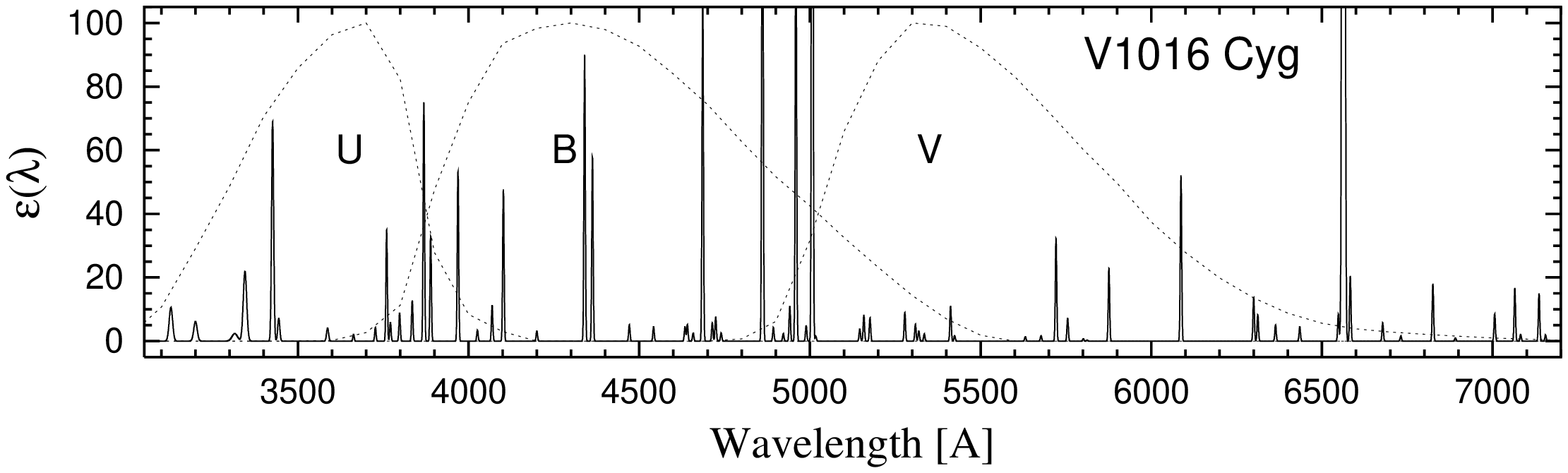,width=37truemm,angle=-90,clip=}}
\captionb{2}{Function $\epsilon (\lambda)$ for the symbiotic nova 
             V1016~Cyg, which represents its emission line spectrum 
             in units of the continuum. Dotted lines show 
             normalized transmission functions 
             of the $U,~B,~V$ filters. 
}
\end{figure}
%
%
\vskip1mm
\begin{figure}
\centerline{\psfig{figure=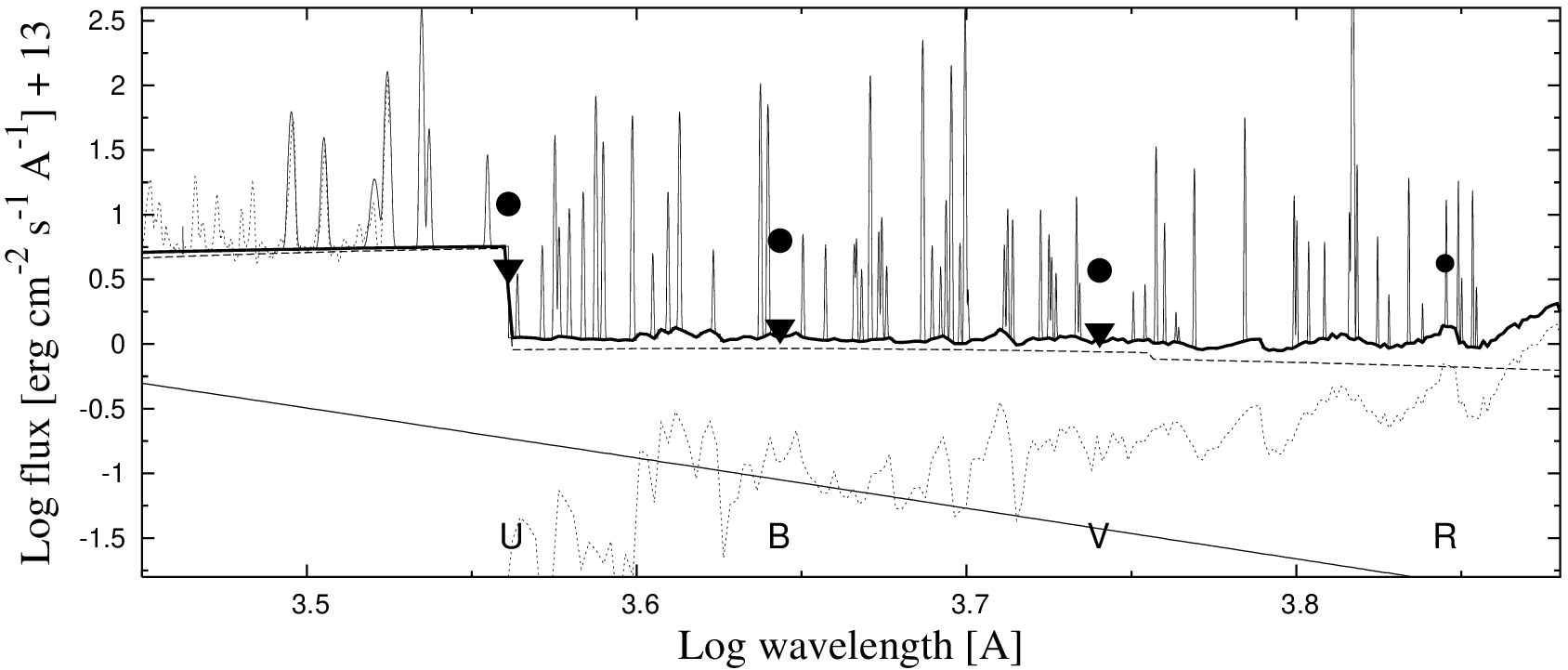,width=52truemm,angle=-90,clip=}}
\captionb{3}{
Reconstructed SED in the near-UV/optical continuum of the symbiotic 
nova V1016~Cyg. Solid thin line represents radiation from 
the hot object ($T_{\rm h}$ = 150000\,K), dashed line is that from 
the nebula ($T_{\rm e}$ = 16800\,K) and the solid thick 
line is the resulting modeled continuum. Radiation from 
the giant was compared with the synthetic spectrum and scaled 
to the flux in the $J$ band (see the text). The emission line 
spectrum was introduced in Fig.~2. Full circles are fluxes given 
by the broad-band photometry. Triangles are the $U,~B,~V$ 
magnitudes corrected for the emission lines.
}
\end{figure}

\sectionb{3}{THE EXAMPLE OF V1016~Cyg}

V1016~Cyg is a symbiotic nova, which erupted in 1964 when 
it brightened by about 5\,mag in the optical and has continued 
a very slow decrease of its brightness (Parimucha et al. 2000). 
It belongs to the D-type symbiotics (strong IR dust emission) 
and contains a Mira variable as cool component 
(SS90 and references therein). According to the ultraviolet 
(IUE) and optical (INT) observations, the near-UV/optical 
spectral region was dominated by the nebular continuum 
superposed with strong emission lines at high 
excitation/ionization degrees (SS90). 

To reconstruct the optical continuum of V1016~Cyg we used
the IUE spectra SWP24655 and LWP04959 taken on 10/12/84 
and the synthetic spectrum for the red giant with
$T_{\rm eff} = 3\,100$\,K and $\log(g) = 0.5$ 
(Hauschildt et al. 1999). The latter was scaled to 
the observed flux in the $J$-band, which is assumed 
to be free of any dust emission. The emission line spectrum 
is shown in Fig.~2 and was described in Sect. 2.4. 
All observations here were dereddened for interstellar 
extinction with $E_{\rm B-V}$ = 0.28 (SS90). 

Results are drawn in Fig.~3. Our model confirms a strong 
contribution from the symbiotic nebula to the optical 
wavelengths. Applying our procedure described in Sect. 2 
we found that the removal of emission lines makes the star's 
brightness fainter by 1.23, 1.67 and 1.18\,mag in 
the $U,~B$ and $V$ band, respectively. This means that V1016~Cyg 
emitted about 68\%, 78\% and 66\% of the total light 
throughout the emission lines in these passbands. 
The corrected fluxes fit perfectly the predicted continuum. 
In addition, the effect is different in different passbands,
which results in the relevant change of the colour indices.
We calculated the empirical indices according to Eqs.~4 and 5. 
By this way we determined quantities 
$(U-B)$ = -1.27, 
$(B-V)$ = +0.03 
for the observed spectrum (i.e. including lines) 
and 
$(U-B)_{\rm cont}$ = -1.68, 
$(B-V)_{\rm cont}$ = +0.53 
for the modelled continuum. 
This results in a rather significant change in the colour indices, 
$\Delta (U-B)$ = 0.41 
and 
$\Delta (B-V)$ = -0.50. 
Therefore, it is necessary to remove the excess due to emission
lines from the $U,~B,~V$ magnitudes before using any diagnostic
by colour indices. 

\vskip7mm

ACKNOWLEDGMENTS. 
This work was supported by Science and Technology Assistance
Agency under the contract No. APVT-20-014402. 
The author thanks prof. V. Strai\v{z}is for some comments. 

\goodbreak

\References
\ref
A\v{z}usienis~A., Strai\v zys~V. 1969, AZh, 46, 402 
\ref
Fern\'andez-Castro~T., Gonz\'alez-Riestra~R., Cassatella~A., 
       Taylor~A.~R., Sea\-quist~E.~R. 1995, ApJ, 442, 366
\ref
Hauschildt~P.~H., Allard~F., Ferguson~J., Baron~E., Alexander~D.~R. 
      1999, ApJ, 525, 871
\ref
Matthews~T.~A., Sandage~A.~R. 1963, ApJ, 138, 30 
\ref
Nussbaumer~H., Vogel~M. 1989, A\&A, 213, 137 
\ref
Parimucha~\v{S}, Arkhipova~V.~P., Chochol~D., Kroll~P., Pribulla~T., 
    Shugarov~S.~Yu., Ulyanikhina~O., Chinarova~L.~L. 2000,
    Contrib. Astron. Obs. Skalnat\'e Pleso, 30, 99
\ref
Schmid~H.~M., Schild~H. 1990, MNRAS, 246, 84
\ref
Skopal~A. 2001, Contrib. Astron. Obs. Skalnat\'e Pleso, 31, 119
\ref
Skopal~A. 2003, A\&A, 401, L17
\ref
\end{document}